\documentclass[12pt]{article}

\title{\bf Rotating Relativistic Thin Disks as Sources of the Taub-NUT
Solution}

\author{Guillermo A. Gonz\'{a}lez\thanks{e-mail: guillego@uis.edu.co}	\\
{\it Escuela de F\'{\i}sica, Universidad Industrial de Santander}	\\
{\it A.A. 678, Bucaramanga, Colombia}}

\date{ }

\begin{document}

\maketitle

\begin{abstract}

Rotating disks with nonzero radial pressure and finite radius are
studied. The models are based in the Taub-NUT metric and constructed
using the well-known ``displace, cut and reflect'' method. We find that
the disks are made of perfect fluids with constant energy density and
pressure. The energy density is negative, but the effective Newotnian
density is possitive as the strong energy condition requires. We also
find that the disks are not stable under radial perturbations and that
there are regions of the disks where the particles move with superluminal
velocities.

\end{abstract}

\section{Introduction}

An important problem in general relativity is the obtention of exact
solutions of Einstein equations corresponding to physically acceptable
configurations of matter. Although exact solutions has been obtained
only in simple, highly symmetric, cases, in the last twenty years several
generation techniques for obtain solutions of Einstein equations from a
given solution have been deve\-loped with success \cite{KSMH}. However,
one of the shortcomings of these methods is that they do not give
significant information about the physical or geome\-trical meaning of
the generated solutions.

As recently has been shown \cite{BLK,BLP}, many vacuum Weyl solutions
can be interpreted as the metrics of static thin disks, constructed
using the well-known ``displace, cut and reflect'' method. The idea of
the method is simple. Given a solution of the vacuum Einstein equations,
a cut is maked above all singularities or sources. The identification
of this solution with its mirror image yields relativistic models of
disks as a consequence of the jump in the normal derivative of the
metric tensor. The disk model can also be used for the interpretation
of vacuum stationary solutions.

In the last years, disks models with radial pressure or tension
\cite{GL1}, electric fields \cite{LZB}, magnetic fields \cite{LET1}
and magnetic and electric fields \cite{KBL} have been studied. See also
the references \cite{BL,PL,GL2}. In this work we apply the above method
to study rotating disks with nonzero radial pressure and of finite
radius. The models are based in the Taub-NUT (Newman-Unti-Tamburino)
metric, which is one of the simplest axially symmetric stationary
solutions of vacuum Einstein equations \cite{KSMH}.

\section{The Taub-NUT Solution}

We can write the metric as the Weyl-Lewis-Papapetrou line element
\cite{KSMH},
\begin{equation}
ds^2 = \ e^{- 2 \Phi} [{\cal R}^2 d\varphi^2 + e^{2 \Lambda} (dr^2 +
dz^2)] \ - \ e^{2 \Phi} (dt + {\cal W} d\varphi)^2 , \label{eq:met}
\end{equation}
where $\Phi$, $\Lambda$, ${\cal W}$ and ${\cal R}$ are functions of $r$
and $z$ only. In order to obtain finite thin disks with nonzero radial
pressure, we take a solution of vacuum Einstein equationes defined by
the following relations \cite{GL1}.

Let be $w = r + i z$ and ${\cal F}
(w) \ = \ w + \alpha \sqrt{w^2 - 1}$, with $\alpha \geq 1$. Then
\begin{eqnarray}
{\cal R} (r,z) \ &=& \ {\rm Re} \ {\cal F} (w) \ , \label{eq:com} \\
&&	\nonumber	\\
{\cal Z} (r,z) \ &=& \ {\rm Im} \ {\cal F} (w) \ , \\
&&	\nonumber	\\
\Phi (r,z) \ &=& \ \Psi ({\cal R},{\cal Z}) \ , \\
&&	\nonumber	\\
\Lambda (r,z) \ &=& \ \Pi ({\cal R},{\cal Z}) \ + \ \ln
|{\cal F}' (w)| \ , \\
&&	\nonumber	\\
{\cal W} (r,z) \ &=& \ {\cal M} ({\cal R},{\cal Z}) \ . 
\end{eqnarray}
Now we take $\Psi$, $\Pi$ and ${\cal M}$
as given by the Taub-NUT solution, that can be written in prolate spheroidal coordinates as \cite{RT}
\begin{eqnarray}
\Psi &=& \frac{1}{2} \ln \left[ \frac{x^2 - 1}{x^2 + 2 u x + 1}
\right] ,      \\
	&	&	\nonumber	\\
\Pi &=& \frac{1}{2} \ln \left[ \frac{x^2 - 1}{x^2 - y^2}
\right] ,	\\
	&	&	\nonumber	\\
{\cal M} & = & 2 k v y .
\end{eqnarray}
The possitive constants $u$ and $v$ can be written as $u \ = \ m/k$ and
$v \ = \ l/k$, with $k^2 = m^2 + l^2$, where $m$ is the mass and $l$ is
the NUT parameter \cite{KSMH}, and thus $u^2 + v^2 = 1$.  The relation
between $(x,y)$ and $({\cal R},{\cal Z})$ is given by
\begin{equation}
{\cal R}^2 = k^2 (x^2 - 1) (1 - y^2) , \qquad {\cal Z} = k x y,
\label{eq:pro}
\end{equation}
where $1 \leq x \leq \infty$, $0 \leq y \leq 1$ and $k = \sqrt{\alpha^2
- 1}$.

\section{The Energy-Momentum Tensor}

The energy-momentum tensor of the disks can be computed by using the
distributional approach, see \cite{GL1,GL2}, and can be written as
\begin{eqnarray}
&S^0_0 &= \ \frac{e^{\Phi - \Lambda}}{{\cal R}^2} \{ 2 {\cal R}^2
(\Lambda,_z - \ 2 \Phi,_z) + \ 2 {\cal R}{\cal R},_z - \ e^{4\Phi}
{\cal W} {\cal W},_z \} , \label{eq:emt1}  \\
&	&	\nonumber	\\
&S^0_1 &= \ \frac{e^{\Phi - \Lambda}}{{\cal R}^2} \{ 2 {\cal R} {\cal W}
({\cal R},_z - \ 2 {\cal R} \Phi,_z) - \ ( {\cal R}^2 + \ {\cal W}^2
e^{4\Phi} ) {\cal W},_z \} , \label{eq:emt2}  \\
&	&	\nonumber	\\
&S^1_0 &= \ \frac{e^{\Phi - \Lambda}}{{\cal R}^2} \{ e^{4\Phi} {\cal
W},_z \}, \label{eq:emt3} \\
&	&	\nonumber	\\
&S^1_1 &= \ \frac{e^{\Phi - \Lambda}}{{\cal R}^2} \{ 2 {\cal R}^2
\Lambda,_z + \ e^{4\Phi} {\cal W} {\cal W},_z \} , \label{eq:emt4} \\
&	&	\nonumber	\\
&S^2_2 &= \ \frac{e^{\Phi - \Lambda}}{{\cal R}^2} \{ 2 {\cal R} {\cal
R},_z \} , \label{eq:emt5}
\end{eqnarray}
where all the quantities are evaluated at $z = 0^+$, $0 \leq r \leq
1$. Is easy to see that, using (\ref{eq:com}) -- (\ref{eq:pro}), the
energy-momentum tensor can be cast as
\begin{equation}
S_{ab} \ = \ ( \sigma + p ) V_a V_b \ + \ p \ h_{ab} \ ,
\end{equation}
where
\begin{eqnarray}
\sigma &=& - \frac{2 p}{\alpha} \left[ \frac{\alpha + k u}{\alpha^2 +
2 u \alpha k + k^2} \right] \ , \\
&&	\nonumber	\\
p &=& \frac{2 \alpha}{\sqrt{\alpha^2 + 2 u \alpha k + k^2}} \ .
\end{eqnarray}
$V^a$ is the velocity vector of the disk, with components
\begin{equation}
V^a \ = \ e^{- \Phi} ( 1, 0 , 0, 0 ) \ ,
\end{equation}
and $h_{ab}$ is the metric of the $z = 0$ hypersurface. The disks so are
made of perfect fluids with constant energy density and pressure. As we
can see, $\sigma \leq 0$, and so the disks do not agree with the weak
energy condition \cite{HE}. On the other hand, the effective Newtonian
density, defined as $\varrho = \sigma + 2 p$, is
\begin{equation}
\varrho \ = \ \frac{2 k p}{\alpha} \left[ \frac{u \alpha^2 + 2 \alpha k +
u k^2}{\alpha^2 + 2 u \alpha k + k^2} \right] \ ,
\end{equation}
so that $\varrho \geq 0$, as the strong energy condition requires. 

\section{The Motion of the Disks}

In order to analize the motion of the disks we compute its tangential
velocity with respect to the locally nonrotating frames \cite{BAR,BPT},
\begin{equation}
{\rm V} = \frac{g_{11} ( \Omega - \omega )}{\sqrt{g_{01}^2 - g_{00}
g_{11}}} \ ,
\end{equation}
were $\omega = - g_{01}/g_{11}$ and $\Omega = V^1/V^0$. By using
(\ref{eq:com}) -- (\ref{eq:pro}) we obtain
\begin{equation}
{\rm V} \ = \ - \left[ \frac{k v p^2}{2 \alpha^2} \right] \frac{\sqrt{1 -
r^2}}{r} .
\end{equation}
As we can see from the above expression, the particles of the disks move with superluminal velocities for $r < r_0$, where
\begin{equation}
r_0 \ = \ \frac{k v p^2}{\sqrt{4 \alpha^2 + k^2 v^2 p^4}} \ .
\end{equation}
The specific angular momentum of a particle of the disk, with mass $\mu$, is given by $h = p_\varphi/\mu = g_{\varphi a} V^a$. Thus we have
\begin{equation}
h^2 \ = \ \left[ \frac{k^2 p^2}{1 + k^2} \right] ( 1 - r^2 ) .
\end{equation}
and is easy to see that
\begin{equation}
\frac{d(h^2)}{dr} \ < 0 \ .
\end{equation}
That is, the disks are not stable under radial perturbations, as can be
concluded by an extension of Rayleigh criteria of stability of a fluid
in rest in a gravitational field; see, for instance \cite{FLU}.

\section{Concluding Remarks}

We do not know of any exact axially symmetric stationary solution
of Einstein equations with the kind of physical properties of the
above model. We find that the disks are made of perfect fluids with
constant energy density and pressure. The energy density is negative,
but the effective Newotnian density is possitive as the strong energy
condition requires. We also find that the disks are not stable under
radial perturbationsand that there are regions of the disks where the
particles move with superluminal velocities.

We are now working in rotating disks models with nonzero radial pressure
based in the Kerr metric. In this case the energy-momentum tensor of
the disks may be not so simple and we can have zones with heat flow. The
inclusion of electric or magnetic fields to these models is also under
consideration in order to obtain ``hot'' rotaing disks, with or without
radial pressure.

\end{document}